\documentclass[12pt]{JHEP}
\usepackage{amssymb}

\newcommand{\Tr}{\mathop{\rm Tr}\nolimits}

\newcommand{\leftpartial}{\mathop{\!\stackrel{\leftarrow}{\partial}}\nolimits}
\newcommand{\rightpartial}{\mathop{\!\stackrel{\rightarrow}{\partial}}\nolimits}

\def\bra#1{\langle #1 |}
\def\ket#1{|#1 \rangle}

\def\ov{\overline}
\def\lp{\left(}
\def\rp{\right)}

\def \be {\begin{equation}}
\def \ee {\end{equation}}
\def \bea {\begin{eqnarray}}
\def \eea {\end{eqnarray}}

\def \zz {{\mathbb Z}}
\def \cc {{\mathbb C}}
\def \rr {{\mathbb R}}
\def \TT {{\mathbb T}}

\def \aa {{\cal A}}
\def \bb {{\cal B}}
\def \mm {{\cal M}}
\def \eee {{\cal E}}
\def \ss {{\cal S}}

\def \nc {noncommutative }

\def \sf  {string field }
\def \sft {string field theory }

\title{Exact solitons on noncommutative tori}
\author{Thomas Krajewski, Martin Schnabl\\
{Scuola Internazionale Superiore di Studi Avanzati} \\
{Via Beirut 4, 34014 Trieste, Italy} \\
INFN, Sezione di Trieste \\
E-mail: {\tt krajew@fm.sissa.it, schnabl@sissa.it}}

\abstract{We construct exact solitons on noncommutative tori for
the type of actions arising from open string field theory. Given
any projector that describes an extremum of the tachyon potential,
we interpret the remaining gauge degrees of freedom as a gauge
theory on the projective module determined by the tachyon.
Whenever this module admits a constant curvature connection, it
solves exactly the equations of motion of the effective string
field theory. We describe in detail such a construction on the
noncommutative tori. Whereas our exact solution relies on the
coupling to a gauge theory, we comment on the construction of
approximate solutions in the absence of gauge fields. }

\preprint{SISSA
22/2001/FM/EP\\ {\tt hep-th/0104090}}

\begin{document}

\section{Introduction}
Since the original conjectures by Sen \cite{Sen:Descent}  on the
emergence of D-branes as solitons in open string field theory, 
there has been much progress in many directions. In particular it was observed
\cite{HKLM} that introducing large $B$-field enables one to
describe D-branes as \nc solitons \cite{GMS, DMR}. Starting with
the open \sft and integrating out all the massive and unphysical
degrees of freedom one obtains some unknown effective action with
certain conjectured properties. These are just enough to construct
lower dimensional D-branes as solitons in the tachyon field. Later
on it was realized \cite{HKL} that the assumption of large
noncommutativity or $B$-field is not actually necessary. Turning
on appropriate gauge field which sets to zero all the covariant
derivatives of the tachyon, one can construct exact \nc solitons.
For a recent review see \cite{Komaba}.

Most of the work on the \nc solitons has been done so far for the
\nc Moyal plane. Its nice property is that the algebra of
functions on the plane with the Moyal star product is isomorphic
to the algebra of compact operators on a separable Hilbert space
through the Weyl quantization formula. This allows one to use
familiar methods of the quantum mechanical harmonic oscillator.
The next to simple and more realistic problem of solitons on the
\nc torus studied in \cite{Schnabl,Bars,Sahraoui,MM} requires
however more involved mathematics of \nc geometry.

The aim of the present paper is to find exact soliton solutions on the torus
for the type of action arising from \sf theory. This problem has been recently
 addressed in a pioneering work \cite{MM}. Our paper completes the latter in
the sense that it provides a general method for constructing solutions of the
problem they raised.

The effective action contains two fields: the tachyon field and
the gauge field which are coupled through a covariant derivative.
The tachyon equation of motion can be satisfied by picking up any
projector (so that the potential energy is minimal) and finding a
compatible connection (thus the covariant derivative of the
tachyon vanishes). It is a fortunate fact that for a given
projector the space of all compatible connections turns out to be
remarkably simple. Besides, among all the compatible connections
one can select those that solve exactly the remaining equations of
motion of the gauge theory. The latter are just the connections
whose curvature is expressed as a linear combination of projectors
entering in a decomposition of the tachyon field. They are related
to the so-called constant curvature connections that have been
extensively studied in the mathematical literature
\cite{cras,connesrieffel,rieffel1,spera,schwarz}. We provide
explicitly such a link in the simplest possible situation, but our
method is general and yields exact solutions in a much broader
context.

The paper is organized as follows. In section 2 we first review
the \nc effective field theory description of the tachyon and
gauge fields. We show in general how one can satisfy all the
equations of motion and determine the space of all compatible
connections. In section 3 we use the bimodule technique to provide
an explicit solution of the gauge theory. In section 4 we show
that the solutions we have found have the right tension to be
interpreted as D-branes and we demonstrate the independence on the
choice of the Seiberg-Witten two form $\Phi$ used in the
description. Section 5 is devoted to an approximate solution in
the absence of gauge fields. This leads us to a projector, known
as Boca's projector, which is related to the particular solution
found on the Moyal plane. The relation between the GMS projector
and Boca's projector in the large torus limit is further studied
in the appendix.

\section{Effective description of string field theory}

We start by considering the open bosonic string theory propagating
in the closed string background of the form ${\cal M} \times
\TT^{d}$ where ${\cal M}$ is arbitrary 26-$d$ dimensional
manifold. The effect of turning on a constant $B$-field (for a
review see \cite{SW}) along a flat submanifold, in this case the
torus $\TT^{d}$, is neatly described by replacing ordinary
products by the \nc star products defined as
\be
f*g= f\, e^{\frac{i}{2}\theta_{Moyal}^{ij} \leftpartial_i
\rightpartial_j} g,
\ee
where $\theta_{Moyal}^{ij}$ is an antisymmetric real matrix.

One also has to replace the ordinary closed string metric $g_{ij}$
and coupling constant $g_s$ by effective open string parameters $G_{ij}$ and
$G_s$.
All these effective parameters are related to the closed ones through
\bea\label{openparam}
\frac{1}{G+2\pi\alpha' \Phi} &=& -\frac{\theta_{Moyal}}{2\pi\alpha'} +
\frac{1}{g+2\pi\alpha' B},
\nonumber\\
G_s &=& g_s \left( \frac{\det(G+2\pi\alpha' \Phi)}
{\det(g+2\pi\alpha' B)} \right)^{\frac 12},
\eea
where $\Phi$ is the Seiberg-Witten two form \cite{SW,Seiberg} which always
appears in the combination with the gauge field strength\footnote{We are using
geometric conventions where $A$ and $F$ are antihermitian. The relation with
\cite{SW} is $A^{SW}=i A^{here}$.} $iF+\Phi$. Its appearance is related to a
freedom in the regularization of the worldsheet theory.

All the fields are regarded as functions of the commutative coordinates $x^a$
and are valued in the \nc algebra $\aa_\theta$. In more concrete terms, this
algebra is generated by functions $U_i = e^{i\frac{x^i}{R}}$ having the
commutation relations
\be
U_i * U_j = e^{-\frac{i}{R^2} \theta_{Moyal}^{ij} } U_j * U_j.
\ee
For simplicity we take all radii of the torus equal to $R$, but all what follows
is easily adapted to any constant metric.

To make contact with the standard conventions for the \nc torus we
shall omit the stars and set
\be
2\pi \theta^{ij} = -\frac{\theta_{Moyal}^{ij}}{R^{2}}.
\ee
The
algebra is spanned by $U_{\vec{n}}=\prod_{i=1}^{d} U_i^{n_i}$
where $\vec{n} \in \zz^{d}$.

The standard partial derivatives with respect to the coordinates $x^{i}$ are
derivations of the algebra $\aa_{\theta}$, i.e. they also satisfy the Leibniz
rule for the star product.

The ordinary integral over the torus yields a trace on the
algebra, i.e. two elements of $\aa_{\theta}$ commute under the
integral. In accordance with the standard trace that can be found
in the mathematical literature, we choose to normalize it as
follows,
\be
\frac{1}{(2\pi R)^{d}} \int_{\TT^{d}} \sum_{\vec{n} \in \zz^{d}}
a_{\vec{n}} U_{\vec{n}} = a_{\vec{0}}.
\ee
Accordingly, this trace will be referred to as the ``normalized
integral''. We refrain from calling it the trace, since one also
considers matrices with entries in $\aa_{\theta}$. On the algebra
$M_{N}(\aa_{\theta})$, one introduces the ordinary linear form
$\Tr$ from $M_N(\aa_{\theta})$ to $\aa_{\theta}$ as the sum of the
diagonal elements. It is not a trace since the property
$\Tr(AB)=\Tr(BA)$ is lost when the algebra is not commutative.
However, this property is true after integration.

The effective action is obtained by integrating out all the fields
from open \sft except the tachyon $T$ and the gauge field $A_\mu$.
It takes the general form
\bea\label{action}
S &=& \frac{c}{G_s} \int_{\cal M} d^{26-d}x \int_{\TT^{d}}
\sqrt{\det G}\, \Tr\left[ \frac 12 f(T^2-1) G^{\mu\nu} D_\mu T
D_\nu T - V(T^2-1) \right.
\nonumber\\
&& \qquad\qquad \left.
-\frac 14 h(T^2-1) (iF_{\mu\nu}+\Phi_{\mu\nu})
(iF^{\mu\nu}+\Phi^{\mu\nu}) + \cdots \right],
\eea
where $\mu,\nu = 1,2,\ldots,26$ and $c=g_s T_{25}$ is a $B$
independent constant. The tachyon field $T$ is a $\aa_{\theta}$-valued
function on ${\cal M}$ and transforms in the adjoint
representation. The gauge field $A_{\mu}$ is an antihermitian
matrix of functions with values in $\aa_{\theta}$ and its
curvature is defined as usual. This action is invariant under the
standard noncommutative gauge transformations.

The explicit form of the effective action is not completely known.
It may contain higher order terms that are represented in
(\ref{action}) by dots. The latter are constructed using products
of higher order covariant derivatives of the tachyon field and of
the curvature tensor. Fortunately their explicit form is not
necessary in order to apply the method we shall describe below.

Furthermore, the functions $f,h,V$ are only required to satisfy
certain conjectured properties. The following ones are not
intended to be the complete list but merely those we shall need in
what follows,
\be
\begin{array}{cclcclc}
V(0)  &=& 0,\quad &  h(0) &=& 0, \quad & f(0)=0, \\
V(-1) &=& 1,\quad & h(-1) &=& 1, \quad &  \\
V'(0) &=& 0,\quad & h'(0) &=& 0. \quad &
\end{array}
\ee
The tachyon potential is normalized in such a way that at the
closed string vacuum $T=1$ it has a minimum equal to zero
and at $T=0$ it has a local maximum equal to $1$. The conditions
for $h(0)$ and  $f(0)$ reflect Sen's conjecture that at the closed
string vacuum all kinetic terms of physical excitations do vanish.
This conjecture has been recently tested numerically
\cite{Hata,ET} and was also a starting point of \cite{RSZ1,RSZ2}.
The condition on $h(-1)$ is just a normalization. What is less
clear is the physical meaning of $h'(0)=0$. It follows
nevertheless from the Dirac-Born-Infeld extension
\cite{Sen:Supersym, Sen:Some} (see also \cite{Bergshoeff, Kluson})
which implies $h =(2\pi\alpha')^2 V$. Among the  higher derivative
terms, those constructed solely from the curvature $F$ and the
tachyon $T$ (without derivatives) will be quite important later.
Because of the Dirac-Born-Infeld extension, their coefficients
$h_{(n)}$ also satisfy $h_{(n)}(0)=0$ and $h_{(n)}'(0)=0$.


The action (\ref{action}) leads to two equations of motion
obtained by variation with respect to tachyon and gauge fields.
They are obtained by expanding the functions $f$, $g$ and $V$ in
power series and collecting the coefficients of $\delta T$ and
$\delta A_{\mu}$ after repeated integrations by parts and cyclic
permutations of the fields. For the sake of brevity, we do not
give their explicit form here. Indeed, they are quite cumbersome
due to the noncommutativity between the fields and their
variations. Instead, we list a few conditions on the fields that
are sufficient in order to solve the equations of motion
\bea\label{eom}
V'(T^2-1) T &=& 0,
\nonumber\\
D_\mu T &=& 0,
\nonumber\\
D_\alpha F_{\mu\nu} &=& 0,
\nonumber\\
\int_{\TT^{d}} \left[ \delta h(T^2-1)  (iF_{\mu\nu}+\Phi_{\mu\nu})
(iF^{\mu\nu}+\Phi^{\mu\nu}) \right] &=& 0,
\nonumber\\
\cdots &&
\nonumber\\
\int_{\TT^{d}} \left[ \delta h_{(n)}(T^2-1)  (iF+\Phi)^{2n}
\right] &=& 0,
\nonumber\\
\cdots &&.
\eea
These conditions imply the equations of motion whatever the higher
order terms are. Indeed, the first one ensures that the
contribution to $\delta T$ arising from the tachyonic potential is
identically zero. After integration by parts, the second and the
third one imply that any contribution containing covariant derivatives of
the fields vanish. Finally the last ones are the remaining
contributions from the monomials in the curvature and the tachyon
field without any covariant derivative.

Thanks to the condition $V'(0)=0$ the first equation is solved by taking $T$
to be any projector which we write as $T=1-P$. Therefore the second equation
becomes
\be\label{DP'}
\partial_{\mu} P + A_{\mu} P - P A_{\mu} =0.
\ee
For later purposes it is convenient to solve a slightly more
complicated equation. Suppose that we are given $N$ hermitian
elements $P_{1},\dots,P_{n}$ of the algebra $\aa_{\theta}$ such
that
\be
P_{i}P_{j}=\delta_{ij}P_{i}\qquad\mathrm{and}\qquad
\mathop{\sum}\limits_{1\leq i\leq n} P_{i}=1.\label{projectors}
\ee
This implies that they form a set of mutually orthogonal  projectors whose sum
is the identity. In particular, if we have a single projector $P$, then $P$ and
$1-P$ form such a set.

The generalization of the single equation (\ref{DP'}) is the set
of equations
\be
dP_{i}+AP_{i}-P_{i}A=0\label{DP}
\ee
whose general solution is given by
\be
A=\mathop{\sum}\limits_{1\leq i\leq n} A_{i}+P_{i}dP_{i}
\ee
where $A_{i}$ are arbitrary hermitian elements of $\aa_{\theta}$
such that $P_{i}A_{i}P_{i}=A_{i}$. The solution is easily obtained
by decomposing $A$ into its ``matrix elements''
\be
A=\mathop{\sum}\limits_{1\leq i,j\leq n} P_{i}A P_{j}.
\ee
Whereas the diagonal terms $P_{i}AP_{i}$ are arbitrary, the
off-diagonal ones are completely fixed by (\ref{DP}). In these formulae, we
have used coordinate the free notations of differential geometry.  The reader
should bear in mind that these are just a convenient notation  that avoids a
proliferation of space-time indices and they are defined in the noncommutative
case by the same formulae as in the commutative case.

Let us suppose that $P$ has been written as
\be
P=\mathop{\sum}\limits_{1\leq i\leq m} P_{i}
\ee
and that
\be
1-P=\mathop{\sum}\limits_{m+1\leq i\leq n} P_{i},
\ee
where $P_{1},\dots,P_{n}$ is a system of mutually orthogonal
projectors whose sum is one. Such a system always exists (take for
instance $P$ and $1-P$), but considering a more general solution
will be useful in the sequel.

Then the gauge field $A$ we have constructed satisfies (\ref{DP}).
Summing over $i$ from 1 to $M$, one concludes that it also
satisfies (\ref{DP'}). Accordingly, we have solved the equation
$DT=0$ and we are left with the last equations involving the
curvature.

The curvature $F=dA+A^{2}$ is easily computed using
\be
dP_{i}P_{j}+P_{i}dP_{j}=\delta_{ij}dP_{i}.
\ee
It turns out to be diagonal
\be
F=\mathop{\sum}\limits_{1\leq i\leq n} \lp
P_{i}dP_{i}dP_{i}P_{i}+P_{i}dA_{i}P_{i}+A_{i}^{2}\rp.\label{curvature}
\ee

This fact is quite crucial since  it allows us to satisfy
identically the last set of equations in (\ref{eom}) for the
tachyon. To see that, let us expand $h$ in power series, bearing
in mind that $h(0)=h'(0)=0$,
\be\label{mon}
h(T^{2}-1)=\mathop{\sum}\limits_{n\geq 2}\frac{1}{n!}h^{(n)}(0)
\lp T^{2}-1 \rp^{n},
\ee
so that its variation is easily computed by varying all monomials.
The variation of a monomial of degree $n$ gives rise to $2n$ terms
\be
\delta\lp T^{2}-1\rp^{n}= \mathop{\sum}\limits_{0\leq k\leq
n-1}\lp T^{2}-1\rp^{k} \lp T\delta T+\delta T T\rp \lp
T^{2}-1\rp^{n-1-k}.
\ee
Although $\delta T$ is completely arbitrary, when $T=1-P$ is a
projector, this formula simplifies into
\be
\delta \lp T^{2}-1\rp^{n}=(-1)^{n} \lp T\delta T \lp
T^{2}-1\rp+\lp T^{2}-1\rp\delta T T \rp
\ee
so that it vanishes identically when multiplied with a diagonal
quantity and integrated. The same analysis remains true for the
higher order terms involving $h_{(n)}$ so that we can conclude
that the last set of equations in (\ref{eom}) is satisfied.

We are thus left with the only remaining equation $D_\alpha
F_{\mu\nu}=0$. A further simplification occurs because of
$D_{\alpha}P=0$. This allows to solve the last equation as
follows. Suppose we can find gauge fields $A_i$ (with
$P_{i}A_{i}P_{i}=A_{i}$) such that
\be
P_{i}dP_{i}dP_{i}+P_{i}dA_{i}P_{i}+A_{i}^{2}= \lambda_{i} P_{i}\label{const}
\ee
with $\lambda_{i}\in\cc$ being numbers. Then the compatibility
condition $(\ref{DP})$ implies
\be
D_{\alpha}F=\mathop{\sum}\limits_{1\leq i\leq n}\lambda_{i}D_{\alpha}P_{i}=0,
\ee
which ensures that the full set of equations of motion are
satisfied, whatever the higher order terms are. The existence of
such gauge fields as well as some explicit constructions will be
given in the next section.

\section{Constant curvature connections}

To solve (\ref{const}) we have to make contact with some available
results in the mathematical literature. In fact, it is equivalent
to the existence of a constant constant curvature connections on
the projective module determined by $P_{i}$. On the noncommutative
tori this has been extensively studied, yet formulated in a rather
different way, using functional analytic methods. This way of presenting 
projective modules over noncommutative tori and the associated constant 
curvature connections have been proposed by A. Connes more than twenty years 
ago \cite{cras}.

To proceed, let us first recall that a projector $P$ in
$M_{N}(\aa_{\theta})$ determines a finitely generated projective
module $\eee=P\aa_{\theta}^{N}$. Alternatively, one can consider
$\eee$ as the subspace of elements $\xi$ of $\aa_{\theta}^{N}$
that fulfill $P\xi=\xi$. This is naturally a right $\aa_{\theta}$
module since it is stable by multiplication by elements of
$\aa_{\theta}$ on the right.

Projective modules are fundamental objects in noncommutative
geometry since they provide a generalization of vector bundles.
Moreover, one can define the analogue of a covariant derivative
and its curvature for these modules. In fact, a covariant
derivative $\nabla_{\mu}$ associated to the partial derivative
$\partial_{\mu}$ is just a linear map from $\eee$ to itself
satisfying the Leibniz rule
\be
\nabla_{\mu}(P\xi f)=\nabla_{\mu}(P\xi) f+P\xi\partial_{\mu} f
\ee
for any $P\xi\in\eee$ and $f\in\aa_{\theta}$. It is not difficult to construct
a covariant derivative just by taking the ordinary partial derivative and
projecting the result onto $\eee$ using $P$. This defines the covariant
derivative $\nabla_{\mu}^{0}$ by
\be
\nabla_{\mu}^{0}(\xi)=P\partial_{\mu}(P\xi)
\ee
for any $P\xi\in\eee$.

To the module $\eee$ we also associate its algebra of endomorphisms
End$(\eee)$ which is the algebra of linear maps from $\eee$ to itself that
commute with the right action of $\aa_{\theta}$. It is isomorphic to the
algebra of matrices of the form $PAP\in M_{N}(\aa_{\theta}) $, or
equivalently, that satisfy $PAP=A$, the sum and product being the ordinary
operations on matrices but the unit is the projector $P$ instead of 1.

It can be easily shown that all connections are of the form
\be
\nabla_{\mu}=\nabla_{\mu}^{0}+A_{\mu},\label{conn}
\ee
where $A_{\mu}$ is a matrix satisfying $PA_{\mu}P=A_{\mu}$. In a
language more familiar to physicists $A_{\mu}$ may be identified
with the gauge field. The gauge transformations are the unitary
elements of End$(\aa_{\theta})$, that is those $g\in$End$(\eee)$
that fulfill $gg^{*}=g^{*}g=P$. Note that one usually imposes that
$A_{\mu}$ be antihermitian and it transforms in the ordinary way
under gauge transformations.

The curvature tensor is defined as the commutator of the covariant derivatives,
\be
F_{\mu\nu}=[\nabla_{\mu},\nabla_{\nu}].
\ee
Unlike the covariant derivatives, it commutes with the left action of
$\aa_{\theta}$, so that it may be identified with an element of
End$(\aa_{\theta})$.

If we compute the curvature of the connection defined in (\ref{conn}), we get
\be
F_{\mu\nu}=P[\partial_{\mu}P,\partial_{\nu}P]P
+P\partial_{\mu}A_{\nu}P-P\partial_{\nu}A_{\mu}P+[A_{\mu},A_{\nu}]
\ee
which has exactly the form of the diagonal elements of
(\ref{curvature}). Since $P$ is the identity of End$(\eee)$, gauge
fields whose curvature is proportional to $P$ are known in the
mathematical literature as ``constant curvature connections''.

Shifting back to the notations we have been using in the previous
section, let us consider our projector $P$ corresponding to the
tachyon field. This projector determines a projective module, but
it has no reason, {\it a priori}, to admit a constant curvature
connection. However, it is known that, in the irrational case
(i.e. when at least one of the entries of the matrix $\theta_{ij}$
is irrational, \cite{rieffel1,schwarz}), any projective module is
isomorphic to a direct sum of projective modules that admit a
constant curvature connection. Unfortunately, this decomposition
and the associated constant curvature connections are not build
using projectors. Thus our last task is to make the connection
between the two points of view as explicit as possible.

Let $S:P\aa_{\theta}^{N}\rightarrow \eee$ be the isomorphism of
$P\aa_{\theta}^{N}$ onto the direct sum
\be
\eee=\mathop{\sum}\limits_{1\leq i\leq
m}\eee_{i},\label{decomposition}
\ee
where the $\eee_{i}$ are
projective modules that admit a constant curvature connection.
This isomorphism induces an isomorphism between the associated
algebra of endomorphisms so that $P$ is mapped to the identity
endomorphism Id$_{\eee}$ of $\eee$.

Consider $\pi_{i}$ the linear endomorphism of $\eee$ that is the
identity on $\eee_{i}$ and that vanishes on the other factors.
Obviously one has
\be
\pi_{i}\pi_{j}=\delta_{ij}\pi_{i}\qquad\mathrm{and}\qquad
\mathop{\sum}\limits_{1\leq i\leq m} \pi_{i}=\mathrm{Id}_{\eee}.
\ee
Accordingly, the inverse images $P_{i}=S^{-1}(\pi_{i})$ determine a system of
$m$ projectors that satisfy (\ref{projectors}). One can perform the same
construction for $1-P$ to add $n-m$ projectors to complete the system of $n$
projectors involved in (\ref{projectors}).

Each of the $n$ projective modules $\eee_{i}$ admits a constant curvature
connection $\nabla_{\mu}^{i}$. Using the isomorphism $S$, this constant
curvature connection induces a constant curvature connection on the projective
module $P_{i}\aa_{\theta}^{N}$ associated to $P_{i}$ which in turn yields a
solution to our equation (\ref{const}), and finally a solution to the full
equation of motion.

In the two dimensional case, this isomorphism can be described
quite explicitly using the techniques developped in \cite{cras}.
 Let us fix an irrational number $\theta\in
]0,1[$ so that the algebra $\aa_{\theta}$ is generated by two
unitary elements $U_{1}$ and $U_{2}$ such that
\be
U_{1}U_{2}=e^{2i\pi\theta}U_{2}U_{1}.
\ee
$\theta$ is the dimensionless parameter related to the ordinary
Moyal deformation parameter through $\theta_{Moyal}=- 2\pi\theta
R^{2}$, $R$ being the radius of the torus.

It is known \cite{rieffel2} that gauge equivalence classes of
projectors in $\aa_{\theta}$, in the two dimensional case and when
$\theta$ is irrational, are parametrized by the values of the
normalized integral. More precisely, if $P$ is a projector, then
there are two integers $p$ and $q$ such that
\be
\frac{1}{(2\pi R)^{2}}\int_{\TT^{2}}P=p+q\theta.\label{trpro}
\ee
Obviously, if two projectors are gauge equivalent, they yield the same
integral, thus the same numbers $p$ and $q$. Conversely, given any two
integers $p$ and $q$ such that $p+q\theta\in [0,1]$, then there is a
projector in $\aa_{\theta}$ such that (\ref{trpro}) holds and any two such
projectors are equivalent. In more physical terms, this means that $p$ and
$q$ parametrize our vacua.

Let us now focus, for simplicity, on the case $p=0$ and $q=1$.
Thus we are working within the gauge equivalence class of the
Powers-Rieffel projector, which is the simplest possible example
of a non trivial projector in $\aa_{\theta}$. We denote by $P$ a
projector within this class, $P\aa_{\theta}$ is then a right
$\aa_{\theta}$-module ($\aa_{\theta}$ acting by multiplication)
which can be equivalently described as follows \cite{cras}.

Let $S(\rr)$ be the space of complex valued function on $\rr$
which decrease fast at infinity. We define a right action of
$\aa_{\theta}$ by
\bea
(\psi U_{1})(t)&=&\psi(t+\theta)\nonumber
\\
(\psi U_{2})(t)&=&e^{-2i\pi t}\psi(t),
\eea
for any function $\psi$. Accordingly , this turns $S(\rr)$ into a
right module which is in fact finitely generated and projective.

Covariant derivatives $\nabla_{1}$ and $\nabla_{2}$ are defined by
\bea
\nabla_{1}\psi(t)&=&-\frac{i}{\theta R}t\psi(t)\nonumber\\
\nabla_{2}\psi(t)&=&\frac{1}{2\pi R}\frac{d}{d t}\psi(t),\label{cov}
\eea
which satisfy the Leibniz rule with respect to the right action of
$\aa_{\theta}$.
The curvature of this connection is constant,
\be
F_{12}=[\nabla_{1},\nabla_{2}]=-\frac{i}{\theta_{Moyal}},
\ee
and we shall comment on the physical meaning of this value in the next
section.

To describe explicitly the isomorphism between the projective
module determined by the projector $P$ and $S(\rr)$, as well as
the pull back of the connection, it is convenient to use the
so-called "bimodule construction".

In general terms, suppose that we are given two algebras
$\aa$ and $\bb$ and a $(\bb,\aa)$-bimodule $\mm$. This means that $\mm$ is a
vector space equipped with a left action of $\bb$ and a right action of
$\aa$ and that these two actions commute, i.e.
\be
(b\psi)a=b(\psi a)
\ee
for any $\psi\in\mm$, $a\in\aa$ and $b\in\bb$. Besides, let us assume that
$\mm$ comes equipped with two scalar products, $(\cdot,\cdot)_{\aa}$ and
$(\cdot,\cdot)_{\bb}$. The first one takes its values in $\aa$ and
is $\aa$-linear whereas the second one takes its values in $\bb$ and is
$\bb$-linear.

We finally assume that they are compatible,
\be
(\psi,\xi)_{\bb}\zeta=\psi(\xi,\zeta)_{\aa} \ee for any elements
$\psi$, $\xi$ and $\zeta$ of $\mm$. Then one easily shows that if
\be
(\psi,\psi)_{\bb}=1,
\ee
then
\be
P=(\psi,\psi)_{\aa}
\ee
is a projector in $\aa$. Indeed,
\bea
P^{2}&=&(\psi,\psi)_{\aa}(\psi,\psi)_{\aa}
\nonumber\\
&=&\lp\psi,\psi(\psi,\psi)_{\aa}\rp_{\aa}
\nonumber\\
&=&\lp\psi,(\psi,\psi)_{\bb}\psi\rp_{\aa}
\nonumber\\ &=&(\psi,\psi)_{\aa}.
\eea
The use of Morita equivalence to construct projectors has been first 
proposed in \cite{rieffel2} and in fact it lead Rieffel to the discovery of 
his projector. It generalizes the partial isometry introduced in \cite{GMS} 
which cannot be applied to the noncommutative torus. Indeed, the
normalized trace of any projector obtained by a partial isometry
should be one, but then the projector is the identity.

Let us now fix $\aa=\aa_{\theta}$,  $\bb=\aa_{-1/\theta}$ and
$\mm=S(\rr)$. The left action of $\aa_{-1/\theta}$ is
\bea
V_{1}f(t)&=&e^{2i\pi t/\theta}f(t)
\\
V_{2}f(t)&=&f(t+1)
\eea
and the two scalar products are defined as
\bea
(\psi,\chi)_{\aa}&=&\theta\mathop{\sum}\limits_{m_{1},m_{2}\in\zz^{2}}
\int_{\rr}{dt}\ov{f(t+m_{1}\theta)}g(t)e^{2i\pi m_{2}t}
U_{1}^{m_{1}}U_{2}^{m_{2}}\label{Aprod}\\
(\psi,\chi)_{\bb}&=&\mathop{\sum}\limits_{n_{1},n_{2}\in\zz^{2}}
\int_{\rr}{dt}f(t-n_{1})\ov{g(t)}e^{\frac{2i\pi n_{2}t}{\theta}}
V_{1}^{n_{1}}V_{2}^{n_{2}}
\eea
The compatibility condition follows from Poisson resummation. Note
that these formulas can be extended to a much broader context
\cite{MM} and \cite{rieffel1}.

Besides, it is known that there exists an element $\psi_{0}$ in
$S(\rr)$ such that $(\psi_{0},\psi_{0})_{\bb}=1$ \cite{rieffel2}.
The construction of such a $\psi_{0}$ is actually a non trivial
task and it is this condition that is at the origin of the strange
choice of functions involved in the Powers-Rieffel projector.
Thanks to the bimodule technique, we end up with a projector
$P_{0}=(\psi_{0},\psi_{0})_{\aa}$ whose trace is $\theta$.

We know that all projectors of trace $\theta$ in $\aa_{\theta}$ are
gauge equivalent. Thus the projector $P$ we started with is related to
$P_{0}$ by
\be
P=uP_{0}u^{*}
\ee
where $u\in\aa$ is unitary, i.e. $u u^*=u^* u =1$. This also
implies that
\be
P=(u\psi_{0},u\psi_{0})_{\aa}.
\ee
Writing $\psi=u\psi_{0}$, this
allows us to define a map $S$ from $P\aa_{\theta}$ to $S(\rr)$ by
\be
S(Pa)=\psi Pa
\ee
for any $Pa\in\aa_{\theta}$. It is a right $\aa_{\theta}$-module
homomorphism which turns out to be invertible. Indeed, its inverse
is given by
\be
S^{-1}(\xi)=(\psi,\xi)_{\aa}.
\ee
Therefore, it establishes a right $\aa_{\theta}$-module homomorphism
between $P\aa_{\theta}$ and $S(\rr)$.

Using the covariant derivatives on $S(\rr)$ defined in (\ref{cov})
we define a covariant derivative $S^{-1}\nabla_{\mu}S$ on the
algebra $\aa$. In terms of gauge fields, we have
\bea
S^{-1}\nabla_{\mu}S(Pa)&=&\lp\psi,\nabla_{\mu}(\psi Pa) \rp_{\aa}
\nonumber\\
&=&\lp\psi,\nabla_{\mu}\psi Pa+(\psi,\psi)_{\aa}\partial(Pa)\rp
\nonumber\\
&=&A_{\mu}Pa+P\partial_{\mu}(Pa),
\eea
with
\be\label{Aexplicit}
A_{\mu}=(\psi,\nabla_{\mu}\psi)_{\aa}(\psi,\psi)_{\aa}.
\ee

It can be checked by direct computation, using the Leibniz rule to
derive the scalar products, that the curvature associated to
$A_{\mu}$ is proportional to $P$,
\be
F_{\mu\nu}=P[\partial_{\mu}P,\partial_{\nu}P]P+P\partial_{\mu}A_{\nu}P
-P\partial_{\nu}A_{\mu}P+[A_{\mu},A_{\nu}]=-\frac{i}{\theta_{Moyal}}P.
\ee
Obviously this construction can be generalized to projectors with
other traces and in higher dimensions. One can thus construct,
with a high level of explicitness, gauge fields compatible with
given projectors that extremize the effective action
(\ref{action}). Note that the above formula (\ref{Aexplicit}) has
already appeared in the physical literature in \cite{Furuuchi} in
the study of \nc instantons.

Let us end this section by a comment on the gauge theoretical
aspect of the procedure we followed. First we have found a non
trivial extremum of the potential which is just the projector
$T_P=1-P$. The latter induces spontaneous symmetry breaking since
it is not invariant under the full gauge group $G$. The subgroup
$H_P$ of all unitary elements $u$ such that $u T_P u^{*}=T_{P}$ is
the unbroken subgroup and the compatibility condition (\ref{DP})
just means that we are only considering a gauge theory with the
little group $H_P$ as gauge group. From this interpretation, it
follows that all projectors are equally good. Indeed, trading $P$
for $uPu^{*}$ is just a gauge transformation and it leaves the
physics invariant. Our truly independent solutions are in fact
equivalence classes of projectors. On the noncommutative tori, the
latter are known to form a discrete set.

\section{D-brane tension}

To verify that the soliton solutions we have found correspond to
D-branes, we can calculate their tension as in
\cite{HKLM,HKL,Komaba}. We shall specialize to the solitons on the
\nc two torus for which we were able to make everything rather
explicit. As they are localized in two directions they should
represent codimension two branes, i.e. D23-branes.

Let us take first for the Seiberg-Witten parameter $\Phi$ the
simplest choice $\Phi=-B=-\frac{1}{\theta_{Moyal}}$ which sets to
zero all the terms in the action containing $iF+\Phi$. The only
contribution to the tension then comes only from the potential
term. Using
\be
\frac{\sqrt{\det G}}{G_s} = \frac{2\pi \alpha'}{g_s
|\theta|_{Moyal}}
\ee
we get for the action evaluated on our solution
\be
S= -\frac{2 \pi \alpha' c}{g_s |\theta|_{Moyal}}  \int_{\TT^{d}} P
\int_{\cal M}  d^{26-d}x \sqrt{\det g_{\cal M}},
\ee
since
\be
V(T^2-1)=V(-P)=V(0)+(V(-1)-V(0))P = P.
\ee
As we have discussed above the projectors on the algebra
$\aa_\theta$ are classified according to their trace which belongs
to $(\zz+\theta \zz) \cap [0,1]$. For a projector $P$ of normalized trace
$\theta m$ we get the tension
\be
T= \frac{2 \pi \alpha' c}{g_s |\theta|_{Moyal}} (2\pi R)^2 \theta
m = (2\pi)^2 \alpha' m T_{25} =
 m T_{23}
\ee
which identifies our soliton to be describing $m$ D23-branes.
Curiously the mathematical theorem tells us that the soliton
number cannot exceed $\frac{1}{\theta}$. Physically it is welcome
since the total energy of the D23-branes should be smaller than
the energy of the original unstable D25-brane.

We may ask what would happen if we had chosen different value for
the parameter $\Phi$. Would we end up with the right tension? The
value we took before was special in the sense that all terms
containing the curvature vanished. If we take different value for
$\Phi$ we have to know all terms in the action which do not
contain derivatives. Fortunately these are precisely those terms
provided by the Dirac-Born-Infeld action \cite{Sen:Supersym,
Sen:Some}
\be
S = \frac{c}{G_s} \int_{\cal M} d^{26-d}x   \int_{\TT^{d}} \left[
-V(T^2-1) \sqrt{\det(G + 2\pi\alpha'(iF + \Phi))} + \cdots \right]
\ee
Clearly the $1-P$ part of $F$ will not contribute. We can use
formula
\be
\sqrt{\det(a+b P)}= \sqrt{\det a}\, (1-P) +  \sqrt{\det(a+b)}\, P
\ee
It is then simple exercise to check using (\ref{openparam}) that
\be
\frac{\sqrt{\det(G + 2\pi\alpha'(\theta_{Moyal}^{-1} +
\Phi))}}{G_s} = \frac{\sqrt{|\det \theta_{Moyal}^{-1}|}}{g_s}
\ee
is actually $\Phi$ independent and when we evaluate the action we
get the same tensions as before.

\section{Approximate solutions without gauge fields and selfdual projectors}

In the previous sections we were solving exactly the equations of
motion by turning on an appropriate gauge field. Obviously one may
ask whether it is possible to satisfy the equations without
turning on the gauge field at all. This should indeed be possible
as suggest the results in CSFT \cite{HK,deMelloKoch,MSZ} and BSFT
\cite{bisft1,bisft2}. In the effective field theory approach we
are pursuing here we are then limited only to approximate
solutions neglecting higher derivatives.

Natural procedure is to first exclude all the derivative terms and
search  for the minima of the potential. As is known \cite{GMS}
nontrivial  solutions of $V'(\phi)=0$ can be constructed using
projectors. The next step is to take into account the kinetic
term, which reduces to
\be
S_{kin}[P]=\frac{1}{\lambda^{2}} \int
P\partial_{\mu}P\partial_{\mu}P\label{kinetic}
\ee
when evaluated on a projector $P$. Note that we have summarized
all relevant information on the function $f$ and the effective
couplings into the ``coupling constant'' $\lambda$.  Whereas all
projectors are ground states of the potential, only a few of them
will minimize the correction given by the kinetic term. Let us
thus try to find the extrema of (\ref{kinetic}) on the space of
all projectors.

This problem has been studied in \cite{DKL}. Let us first derive the
equation of motion. If $P$ is a projector, then $P+\delta P$ is a
projector (at the first order) iff $\delta P=[\delta a,P]$ with
$\delta a\in\aa$. Accordingly, the equation of motion are
\be
P\,\Delta P-\Delta P\,P=0 \label{equation} \ee where $\Delta$ is
the standard laplacian. It is worthwhile to notice that this
equation takes a particular form because we search for the extrema
of (\ref{kinetic}) on the space of projectors. If we were instead
working on the space of all scalar fields, the equations of motion
would be completely different and would include also a nontrivial
dependence on the function $f$ appearing in the effective action
(\ref{action}). To our knowledge it is not clear that the extrema
of the full action (in the absence of gauge field and neglecting
the higher order derivatives of the tachyon) can be obtained by
first searching for the zeros of the potential and then minimizing
the kinetic term on these zeros.

However, the problem of minimizing the kinetic term on the space
of projectors is an interesting question since it involves some
topologically stable solutions in two dimension. In fact this
action admits a topological bound
\be
\frac{1}{\lambda^{2}}\int P\partial_{\mu}P\partial_{\mu}P \geq
\frac{1}{\lambda^{2}}\left|i\epsilon_{\mu\nu}\, \int
P\partial_{\mu}P\partial_{\nu}P\right|.\label{inequality}
\ee
This is easily derived from the relation
\be
\int (\partial P P)^{*}\partial P P= \int
P\partial_{\mu}P\partial_{\mu}P - i\epsilon_{\mu\nu}\, \int
P\partial_{\mu}P\partial_{\nu}P,\label{equality}
\ee
where
$\partial=\partial_{1}-i\partial_{2}$ and from the corresponding
equality involving $\ov{\partial}$.

This inequality (\ref{inequality}) is similar to the inequality
arising in four dimensional Yang-Mills theory and is interpreted
as follows. The space of projectors is not connected and on each
of its connected components the action is bound by the LHS.

It is an easy exercise to show that the LHS is invariant under a
small deformation of the projector. In fact, when $\theta$ is
irrational, two projectors lie in the same connected component iff
they have the same trace \cite{rieffel2}. Furthermore, if the
normalized trace of $P$ is $p+q\theta$, then
\be
\frac{\epsilon_{\mu\nu}}{(2\pi R)^{2}}\,
\int P\partial_{\mu}P\partial_{\nu}P=2i\pi
q,\label{top}
\ee
so that $q$ is an analogue of a two dimensional instanton number.
Indeed, it has been obtained in \cite{DKL} as a topological bound in
the study of a noncommutative generalization of a non-linear
$\sigma$-model. It is part of a general theory that encompasses
both this model and the ordinary non-linear field theory with
values in $S^{2}$. Because $S^{2}$ is homeomorphic to the space of
rank one projectors in $M_{2}(\cc)$, it is easy to write the
kinetic term of the standard non-linear $\sigma$ model and
(\ref{top}) is nothing but the winding number of the corresponding
map from $S^{2}$ into itself. This means that it measures the
homotopy class of this map, i.e. it is an element of
$\pi_{2}(S^{2})$.

In the context of noncommutative geometry, the r\^ole of the
homotopy groups is played by the K-theory of the algebra, (i.e.
classes of projectors and unitary elements of matrix algebra over
the algebra of coordinates). They provide, when suitably
differentiated and integrated, quantities that are stable under
small deformations. In a more abstract language, this is
formulated through the pairing of the cyclic cohomology of the
algebra (i.e. all the possible ways to differentiate and integrate
in a ``suitable way'' elements of the algebra) with its K-theory
(i.e. noncommutative analogues of homotopy classes of vector
bundles  and gauge transformation). This theory is fully developed
in the treatise \cite{Connes} and the recent review \cite{Y2K}.

Turning back to the problem of minimizing (\ref{kinetic}), it
follows from (\ref{equality}) that the bound will be saturated iff
$\partial P\, P=0$ (antiself-duality equation) or $\ov{\partial}
P\,P=0$ (self-duality equation). Because of the nonlinear
constraint arising form the fact that $P$ must be a projector,
these equations are not easy to solve.

If $P$ is a self-dual projector, then $1-P$ is an anti self dual
projector, so that it is sufficient to solve the latter.
Fortunately, this non-linear problem can be turned into a linear
one using the bimodule technique.

Using the notation of section 3, we recall that it allows to
construct a projector in $\aa$ provided we have an element
$\psi\in\mm$ such that $(\psi,\psi)_{\bb}=1$. For simplicity, we
restrict our discussion to the homotopy class of the
Powers-Rieffel projector, so that we know that all projectors in
this class are obtained through unit vector in the bimodule ${\cal
N} =\ss(\rr)$. It follows that the resulting projector will
satisfy
\be
\frac{1}{(2\pi R)^{2}} \int P=\theta,\qquad\mathrm{and}
\qquad \frac{\epsilon_{\mu\nu}}{(2\pi R)^{2}}\int
P\partial_{\mu}P\partial_{\nu}P=2i\pi.
\ee
Note that the map from from unit vectors to projectors is not one to one; two
unit vectors yield the same projector iff they differ by a gauge
transformation. Indeed, if $\psi$ and $\chi$ are unit vectors such
that
\be
P=(\psi,\psi)_{\aa}=(\chi,\chi)_{\aa}
\ee
then
\be
\psi=(\psi,\chi)_{\bb}\,\psi\qquad\mathrm{and}\qquad\chi=
(\chi,\psi)_{\bb}\,\chi
\ee
and the gauge transformation is $u=(\chi,\psi)_{\bb}$ which
belongs to the gauge group (i.e. the group of unitary elements of
$\bb$).

A natural way to construct a unit vector $\psi$ is to start with
an arbitrary element $\chi\in\eee$ whose norm $(\chi,\chi)_{\bb}$ is
invertible. Then standard mathematical techniques (holomorphic
functional calculus, for instance) allow us to define the square
root of $(\chi,\chi)_{\bb}$ so that
\be
\psi=\lp (\chi,\chi)_{\bb}\rp^{-1/2}\chi \ee is a unit vector.

The main difficulty of this approach is to determine whether the
norm $(\chi,\chi)$ is invertible or not. For instance, if $\theta>1$, we know
that $(\chi,\chi)_{\bb}$ is not invertible, otherwise we would have
constructed a projector of trace $\theta>1$ in $\aa_{\theta}$,
which is impossible. On the other hand, for $\theta=1/n$, $\bb$ is commutative
and $(\chi,\chi)_{\bb}$ is invertible iff this function does not vanish.

In trading the projectors for the vectors $\chi\in\eee$, we have introduced
spurious gauge degrees of freedom. In fact, two such vectors yield the same
projector iff they differ by a complex (not necessarily unitary)
gauge transformation. By definition such an element belongs to the
group of invertible elements of $\bb$. Therefore, we have to identify the
vectors that yield anti self-dual projectors but differ only by a complex
gauge transformation.

Let us introduce the complex covariant derivatives
\be
\ov{\nabla}=\nabla_{1}+i\nabla_{2}
\ee
associated to the covariant derivatives $\nabla_{1}$ and $\nabla_{2}$
introduced in section 3.

Now we have all the tools to solve the anti self-duality equation.
In fact the projector $P$ associated to a vector $\chi$ of
invertible norm satisfies the equation
\be
\ov{\partial}P P=0
\ee
iff there is $\rho\in \bb$ such that
\be
\ov{\nabla}\chi=\rho\chi.
\ee
This statement follows from an explicit computation of $\ov{\partial} P\,P $ in terms of $\chi$.

Consequently, our problem has been reduced to the following linear
problem. Given any $\rho\in \bb$, we have to find all $\chi\in\eee$ such that
\be
\ov{\nabla}\chi=\rho\chi. \ee
This is a linear equation in $\psi$ and all the powerful methods of linear
functional analysis can be applied to this problem.

The complex gauge group acts on $\rho$ as
\be
\rho\,\longrightarrow\,g\rho g^{-1}+g\partial g^{-1}
\ee
so that $\rho$ is nothing but a complex gauge field. Moreover, if
$\lambda\in \bb$ can be deduced from $\rho$ by a complex gauge
transformation
\be
\lambda=g^{-1}\rho g+g^{-1}\ov{\partial}g
\ee
then $\chi$ satisfies the linear problem associated to $\rho$ iff $g\chi$
satisfies the linear problem associated to $\lambda$. This means
that we have to solve the linear problem only on the orbit of the
action of the complex gauge group on the complex connection. In
other words, if we find a subset of $\bb$ that
intersects each orbit at least once, it is sufficient to solve
linear problem for those values of $\rho$.

The general problem of the study of these orbit spaces is a rather
difficult problem which is tantamount to the study of the moduli
spaces of complex vector bundles over the noncommutative torus. In
the special case we are considering here, one can show that
constants intersect each orbit at least once, thus we have to
solve the equation
\be
\ov{\nabla}f=\frac{\lambda}{2\pi R}f,\label{differential}
\ee
with $\lambda\in\cc$. This is very easy because (\ref{differential}) is
just the differential equation
\be
f^{'}(t)+\frac{2\pi t}{\theta}f(t)-\lambda i f(t)=0,
\ee
whose solution is the gaussian
\be
f_\lambda(t)= A e^{-\frac{\pi t^{2}}{\theta}+\lambda i t}
\ee
$A$ being an arbitrary constant, which is absolutely inessential
since it cancels in the expression for the projector. Projector
based on this function has been for the first time constructed by
Boca \cite{Boca}. The anti self-duality property was recognized in
\cite{DKL}. It has been also studied in \cite{MM}.

Besides, one can show that two values of $\lambda$ differ by a
gauge transformation iff they belong to the same class in
$\cc/(\zz+\tau\zz)$, where $\tau$ is the modular parameter of our
initial torus \cite{DKL}.  Here we have set $\tau=i$, but it is an
easy exercise to work with a general value of $\tau$ and observe
the covariance under transformation in $SL_{2}(\tau)$. Therefore,
the torus parametrizes all instantons in the homotopy class of the
Powers-Rieffel projector.

\bigskip

{\bf Note added:} After this work has been completed we have
received a preprint \cite{Gopakumar} which also deals with
solitons on \nc tori.

\section*{Acknowledgements}
We would like to thank Lubo\v{s} Motl for some discussion and
Prof. Loriano Bonora for his encouragement and reading the
manuscript. Special thanks are also due to Mario Salizzoni who
provided decisive technical help.

\appendix

\section{Moyal plane limit of Boca's projector}

The Powers-Rieffel projector \cite{rieffel2} discussed in the
context of tachyon condensation in \cite{Schnabl,Bars,Sahraoui,MM}
is fully legitimate one in the sense that one can find exact
solitons based on this projector. Nonetheless one may wish to have
projector which does reduce to the nice GMS solitons \cite{GMS} in
the large torus limit keeping $\theta_{Moyal}$ fixed. The
Powers-Rieffel projector is also not suitable for orbifolding
\cite{MM}.

A projector which overcomes these difficulties was constructed by
Boca \cite{Boca} using the bimodule technique starting from the
gaussian function $f=e^{-\frac{\pi t^2}{\theta}}$.  He has proved
for certain range of $\theta$ that $b=\langle f, f \rangle_\bb $
is invertible. Then one can easily check that
\be
P_\theta =\langle b^{- \frac 12} f ,  b^{- \frac 12} f \rangle_\aa
\ee
is a projector. For special values of $\theta=
1/q, q \in \zz$ the projector can be expressed explicitly in terms
of Jacobi's theta functions. For example for $q$ even one has
\be\label{Boca}
P_{\frac 1q} = \frac{ \sum_{r,s=0}^{q-1}
e^{-\frac{\pi i r s}{q}}
\vartheta_{\frac{s}{q},\frac{r}{2}}^{(q)}(U_2,\frac{iq}{2})
\vartheta_{\frac{r}{q},\frac{s}{2}}^{(q)}(U_1,\frac{iq}{2})} {q
\vartheta(U_2^q,\frac{iq}{2}) \vartheta(U_1^q,\frac{iq}{2})},
\ee
where
\bea \vartheta_{\frac{a}{N},b}^{(N)}(U,\tau) &=& \sum_m
e^{\pi i \tau (m+\frac{a}{N})^2 + 2\pi i(m+\frac{a}{N})b}
U^{Nm+a},
\nonumber\\
\vartheta(U,\tau) &=& \sum_m e^{\pi i \tau
m^2} U^m.
\eea
Note that the above formula (\ref{Boca}) makes sense as it stands
since the denominator turns out to be central element in the
algebra. One can translate back this expression in the language of
ordinary functions and the Moyal star product. With some effort
one can check that in the large torus limit keeping
$\theta_{Moyal}$ fixed (which is equivalent to  $q \to \infty$),
the above projector goes to the basic GMS soliton
\be
P=\ket{0}\bra{0} \simeq 2
e^{-\frac{x_1^2+x_2^2}{|\theta_{Moyal}|}}.
\ee
In this limit the theta functions in the denominator do not
contribute. In the numerator the sums over $r$ and $s$ factorize,
after some rearrangement one can replace them again by Jacobi's
theta function, use its duality property and obtain the result
with all the factors right.

An alternative way is to start with the formula for $b^{-1/2} f$
obtained in \cite{MM}. For small $\theta$ it simply reduces to
\be
b^{-1/2} f \sim \sqrt[4]{\frac{2}{\theta}} f.
\ee
From the formula (\ref{Aprod}) we first calculate
\bea
\sqrt{\frac{2}{\theta}} \langle f ,
f \rangle_\aa &=& \theta \sum_{m,n}
e^{-\frac{\pi\theta}{2}(m^2-2imn+n^2)} e^{im \frac{y}{R}} * e^{in
\frac{x}{R}}
\nonumber\\
&=& \theta \,
\vartheta\left(\frac{x}{2\pi R},\frac{i\theta}{2}\right)
\vartheta\left(\frac{y}{2\pi R},\frac{i\theta}{2}\right).
\eea
Note that in the second line the product between the theta
functions is the ordinary commutative one. The theta function is
the basic one defined by
\be
\vartheta(\nu,\tau) = \sum_m e^{\pi i \tau m^2 +2\pi i m \nu}
\ee
with the familiar modular transformation property
\be
\vartheta\left(\frac{\nu}{\tau},-\frac{1}{\tau}\right)=
(-i\tau)^{\frac{1}{2}} e^{\pi i \nu^2 \tau} \vartheta(\nu,\tau).
\ee
From that simply follows the asymptotic behavior
\be
\langle b^{-\frac{1}{2}} f ,  b^{-\frac{1}{2}} f \rangle_\aa \sim
2 e^{-\frac{x_1^2+x_2^2}{|\theta_{Moyal}|}} \ee which is just as
for the GMS soliton.

\end{document}